\documentclass{aa}

\usepackage{graphicx}
\usepackage{rotating}
\usepackage{amsmath}
\usepackage{txfonts}
\usepackage{natbib}
\bibpunct{(}{)}{;}{a}{}{,}

\begin{document}

\authorrunning{A. Galli \& D. Guetta}
\titlerunning{GRB high energy from internal shocks}

\title{Gamma-ray burst high energy emission from internal shocks}

\author{A. Galli\inst{1,2}
        \and D. Guetta\inst{3}
}

\institute{
Istituto Astrofisica Spaziale e Fisica Cosmica - Sezione di Roma, INAF, via del Fosso del Cavaliere, 100-00113 Roma, Italy
\and INFN of Trieste, Padriciano 99, 34012, Trieste, Italy
\and INAF-Osservatorio Astronomico di Roma, via Frascati 33, 00040 Monteporzio Catone, Italy
}

\offprints{alessandra.galli@iasf-roma.inaf.it}
\date{Received ....; accepted ....}
\markboth{ }{}
\abstract{In this paper we study synchrotron and synchrotron self Compton (SSC) emission from internal shocks (IS) during the prompt and X-ray flare phases of gamma-ray bursts (GRBs). The aim is to test the IS model for the flare emission and for whether GRBs can be GeV sources.We determine the parameters for which the IS model can account for the observed prompt and X-ray flares emission, and study the detectability of the high energy SSC emission by the AGILE and GLAST satellites.We find that the detectability of the SSC emission during the prompt phase of GRBs improves for higher values of the fireball Lorentz factor $\Gamma$ and of the temporal variability $t_v$. If IS is the mechanism responsible for the flare emission, and the Lorentz factor of the shells producing the flare is $\Gamma \sim$ 100, the flare light curves are expected to present some substructures with temporal variability $t_v=10-100$ ms which are much smaller than the average duration of flares, and similar to those observed during the prompt phase of GRBs. If one assumes lower Lorentz factors, such as $\Gamma \sim 10 \div 25$, then a larger temporal variability $t_v \sim$ 40 s can also account for the observed flare properties. However in this case we predict that X-ray flares do not have a counterpart at very high energies (MeV-GeV). An investigation on the substructures of the X-ray flare light curves, and simultaneous X-ray and high energy observations, will allow us to corroborate the hypothesis that late IS are responsible for the X-ray flares.

}

\maketitle

\section{Introduction}
\label{intro}

Observations with the \emph{Swift} satellite have shown that the phase of prompt-to-afterglow transition is characterized by a very interesting phenomenology. Typically the tail of the prompt emission, described by a very steep power law segment (temporal decay index $\delta \sim 3-5$), is followed hundreds to thousands of seconds later by a flat power law segment ($\delta \sim~0.5$), and then around tens of thousands of seconds later by the usual afterglow power law emission ($\delta \sim 1.3$). Sometimes there is a third break where the afterglow light curve steepens to a power law $\delta \sim 2$ \citep{nousek06,zhang05}. 
It has been observed that the tail of the prompt emission, as well as the flat power law phase and the third segment are characterized by the presence of flares \citep{chinca07,falcone07}. Flares appearing hundreds of seconds after the burst were observed for the first time by BeppoSAX in a few events \citep[e.g \object{GRB011121}, \object{GRB011211}, \object{XRF011030},][]{piro05, galli06}. Later \emph{Swift}, thanks to its fast re-pointing capabilities, found that flares are a very common phenomenon in GRBs, as they are detectable in $\sim 30-40 \%$ of its bursts sample \citep{obrien06}. Moreover flares are present both in long and short GRBs \citep[e.g. \object{GRB050724},][]{barthelmy05}, in classical long GRB, X-ray rich (XRR) and X-ray flash (XRF) \citep[e.g. \object{XRF050406},][]{romano06}, and in low redshift \citep[e.g. \object{GRB050803},][]{bloom05} and high redshift \citep[e.g. \object{GRB050904},][]{cusumano07,gendre07} events. Several bursts show multiple flares in their X-ray light curves \citep[e.g. \object{GRB050607} and \object{GRB 050730},][]{pagani06,perri07} while others have only one big flare \citep[see e.g. GRB060526,][]{chincarini06}. 

The flare phenomenon is complex, and despite the large number of ideas put forward in literature to explain its origin, no model has completely interpreted the phenomenology of observed flares. The most important difference between models is the duration of central engine activity: some of them require a long-lived central engine, e.g. models involving late internal shocks (LIS) \citep{burrows05,wu05} and delayed external shocks (DES) \citep{piro05,galli06,galli07} produced by a long duration and/or re-activation of the activity of the engine, while others do not require a long duration central engine activity, e.g. models involving LIS from a short duration central engine \citep{zhang05,wu05}, refreshed shocks \citep{rees98,kumar00refresh}, two components jet \citep{meszaros01,lipunov01}, patchy jet \citep{kumar00}, forward shock (FS)-reverse shock (RS) \citep{fan05}, external shock (ES) with a clumpy medium \citep{dermer00,dermer07} and delayed magnetic dissipation in strongly magnetized ejecta caused by external shock \citep{giannios06}.

A temporal and spectral analysis of the first survey of X-ray flares observed by \emph{Swift} has recently been carried out \citep[][]{chinca07,falcone07}. These authors analyzed the first 110 bursts observed by \emph{Swift} and found that 33 GRBs have significant flares in their X-ray light curves, for a total of $\sim 70$ flares. The temporal analysis revealed that there is a positive correlation between the time of flare appearance and its duration, and that flares can be very energetic in comparison with the underlying afterglow emission. \citet{chinca07} find that a correlation exists between the ratio of intensity of successive prompt $\gamma$-ray pulses and that between successive X-ray flares in the same burst. This is an indication that they may be caused by the same mechanism. \citet{falcone07} used several spectral models to fit each flare. Usually flares can be fitted both by a simple absorbed power law or by a Band model \citep{band93}. Some flares have a soft spectrum consistent with that of the afterglow emission \citep[e.g. \object{GRB050712},][]{depasquale06}, however the majority present a hard-to-soft spectral evolution. A spectral evolution of the order of 0.5-1.0 (depending on the index of the electron population) can be explained in the context of the ES if the typical emission frequency is crossing the observational band, and models such as DES \citep{galli06,galli07}, ES on a clumpy medium \citep{dermer07} or refreshed shocks \citep{guetta07} can be applied to some flares of the sample. However according to the studies performed by \citet{chinca07} and \citet{falcone07}, a large fraction of flares could be explained by LIS produced by a long duration central engine.

In this paper we focus on the possibility that flares are produced by LIS and study the high energy emission from flares in this scenario. We plan to extend our study to other models in a future study. In the LIS model the mechanism responsible for the flare emission is the synchrotron radiation from relativistic electrons accelerated in the shocks. An additional radiation mechanism that may also play an important role is the synchrotron self Compton (SSC) emission, i.e. the up-scattering of the synchrotron photons by the same relativistic electrons to much higher energies. The synchrotron and SSC components from IS have been considered in the context of the prompt emission by \citet{guetta03}. Following \citet{guetta03}, we estimate in this work the high energy emission produced by SSC of the X-ray flare photons. 
As noted by several authors, e.g. \citet{wang06}, \citet{fan07} and \citet{galli07}, the detection of the predicted high energy (MeV to GeV) flare emission by observations with AGILE and GLAST combined with the X-ray flares detection by \emph{Swift} would enable us to check the validity of different models proposed to explain the afterglow phenomenology of GRBs.

The paper is organized as follows. In Sect. 2 we report the AGILE and GLAST sensitivities. In Sect. 3 we summarize the calculations of the synchrotron and SSC spectra in \citet{guetta03} for the prompt emission and provide the expression for the peak energy and the flux normalization. In Sect. 4 we show how the peak energies and flux normalization change for X-ray flares and estimate the high energy flare fluxes. Finally in Sect. 5 we give our summary and conclusions.

\section{Detector sensitivities}
\label{detector}

In this section we present an estimate of the sensitivity of the high energy detectors aboard the AGILE and GLAST satellites. This is necessary to assess the detectability of high energy emission during the prompt and flare phases of GRB we calculate in the following sections.

The AGILE satellite is equipped with the GRID (Gamma-Ray Imaging Detector) instrument operating between $30~MeV$ and $50~GeV$. We estimate the GRID sensitivity by adopting the criterion that a detection is made when at least 5 photons are collected by the detector \citep{zhang01}. According to \citet{galli07} when the detector is source dominated, the detection threshold is given by:

\begin{equation}
\label{soglia}
F_{th}(E)=\frac{5}{A_{eff}(E)T_{int}}~cm^{-2}s^{-1}
\end{equation}

where $E$ is the photon energy, $A_{eff}$ is the effective area of the detector at this energy, and $T_{int}$ is the detector integration time. For $T_{int}$=50 s the GRID is source dominated, thus we estimate the GRID sensitivity using the equation above. In calculating the GRID sensitivity we assume its effective area to be $A_{eff}=550~cm^{2}$ throughout the energy range.

The Large Area Telescope (LAT) aboard GLAST (Gamma-ray Large Area Telescope) observes between 30 MeV and 300 GeV.  The LAT effective area varies with energy in all the band; we take the value of the effective area at different energies from the LAT web page, http://www-glast.slac.stanford.edu/software/IS/glast\_lat\_performance.htm. As shown by \citet{galli07} for an integration time $T_{int}=50$ s, the LAT is source dominated at all energies, thus we can estimate its sensitivity using Eq. \ref{soglia}.

\section{Synchrotron and SSC prompt emission from internal shocks}
\label{prompt}

In this section we estimate the synchrotron and SSC emission in the framework of the IS model, following the prescriptions presented in \citet{guetta03}. In this model the flow Lorentz factor $\Gamma$ is assumed to vary on a typical time scale $t_v$ and with an amplitude $\delta \Gamma \sim \Gamma$. The shells collide at a radius $R \approx 2\Gamma^2ct_v=6 \times 10^{13} \Gamma^2_{2.5}t_{v,-2}$ cm, where $\Gamma_{2.5}=\Gamma/10^{2.5}$ and $t_{v,-2}=t_v/(10^{-2}s)$. The internal energy released in each collision is distributed among electrons, magnetic field and protons with fractions $\epsilon_e$, $\epsilon_B$ and $(1-\epsilon_e)$ respectively. The electrons are accelerated in the shocks to a power law distribution of energy $N(\gamma) \propto \gamma^{-p}$, and radiatively cool by the combination of synchrotron and SSC processes, the timescales of which are $t_{syn} \sim 6 \pi m_ec/\sigma_T B^{2}\gamma$ and $t_{SSC}=t_{syn}/Y$, the combined cooling time being $t_c=(1/t_{syn}+1/t_{SSC})^{-1}$=$t_{syn}/(1+Y)$, where $B$ is the magnetic field, and $Y$ is the Compton y-parameter \citep{sari96}$, Y \approx \epsilon_e/\epsilon_B$ for $\epsilon_e << \epsilon_B$ and $Y \approx (\epsilon_e/\epsilon_B)^{1/2}$ for $\epsilon_e >> \epsilon_B$ .The synchrotron spectrum is

\begin{equation}
\label{Fnu_syn_gc<1}
{\nu F_{\nu}\over\nu_{m}F_{\nu_m}}=
\begin{cases}
\Big( \frac{\nu_{sa}}{\nu_{m}} \Big)^{1/2} \Big( \frac{\nu_{ac}}{\nu_{sa}} \Big)^{19/8} \Big( \frac{\nu}{\nu_{ac}} \Big)^3 & \nu<\nu_{ac}\\
\Big( \frac{\nu_{sa}}{\nu_{m}} \Big)^{1/2} \Big( \frac{\nu}{\nu_{sa}} \Big)^{19/8} & \nu_{ac}<\nu<\nu_{sa} \\
\Big( \frac{\nu}{\nu_{m}} \Big)^{1/2} & \nu_{sa}<\nu<\nu_{m} \\
\Big( \frac{\nu}{\nu_{m}} \Big)^{(2-p)/2} & \nu_{m}<\nu<\nu_{M}.
\end{cases}
\end{equation}

where $\nu_{sa}$, $\nu_{ac}$ and $\nu_M$ are defined in \citet{guetta03} and the synchrotron peak energy is given by:

\begin{equation}
\label{piccosincr}
E_p=h \nu_m = 0.12 (1+z)^{-1}(1+Y)^{-1/3} \epsilon_e^{3/2}\epsilon_B^{1/2}L_{52}^{1/2}\Gamma_{2.5}^{-2}t_{v,-2}^{-1}~MeV
\end{equation}

for $p=2.5$. From the normalization condition we find the flux at the peak of the spectrum to be:

\begin{equation}
\nu_mF_{\nu_m}=\frac{L_{iso}}{24 \pi D^2}=1.3 \times 10^{-6} L_{52} D_{28}^{-2}~erg~cm^{-2}s^{-1}
\end{equation} 

where $D_{28}$ is the burst distance in unity of $10^{28}$ cm. 

The SSC spectrum is given by

\begin{equation}
\label{Fnu_SSC}
{\nu F_{\nu}^{SC}\over Y\nu_{m}F_{\nu_m}}=
\begin{cases}
\Big( \frac{\nu_{sa}^{SC}}{\nu_{m}^{SC}} \Big)^{1/2} \Big( \frac{\nu}{\nu_{sa}^{SC}} \Big)^{2} & \nu<\nu_{sa}^{SC} \\
\Big( \frac{\nu}{\nu_{m}^{SC}} \Big)^{1/2} & \nu_{sa}^{SC}<\nu<\nu_{m}^{SC} \\
\Big( \frac{\nu}{\nu_{m}^{SC}} \Big)^{(2-p)/2} & \nu_{m}^{SC}<\nu<\nu_{KN}^{SC}\\
\Big( \frac{\nu_{KN}^{SC}}{\nu_{m}^{SC}} \Big)^{2-p\over 2}\Big( \frac{\nu}{\nu_{KN}^{SC}} \Big)^{1-2p\over 2} & \nu_{KN}^{SC}<\nu<\nu_{M}^{SC}.
\end{cases}
\end{equation}

where $\nu_{sa}^{SC}$, $\nu_M^{SC}$ and $\nu_{KN}^{SC}$ are defined in \citet{guetta03} and the SSC peak energy is:

\begin{equation}
\label{piccoSSC}
E_p^{SC}=h \nu_m^{SC}= 4.6 \times 10^4 (1+z)^{-1} \epsilon_e^{7/2}\epsilon_B^{1/2}L_{52}^{1/2}\Gamma_{2.5}^{-2}t_{v,-2}^{-1}~MeV
\end{equation}
 
for $p=2.5$.

For details about the spectrum at frequencies higher than the Klein-Nishina frequency $\nu_{KN}^{SC}$, see \citet{guetta03bis}. In our estimate of high energy emission we also take into account the suppression of the emission due to pair production \citep{guetta03}.

\begin{figure}[!htb]
\centering
\includegraphics[width=7.0cm,height=9.16cm,angle=90]{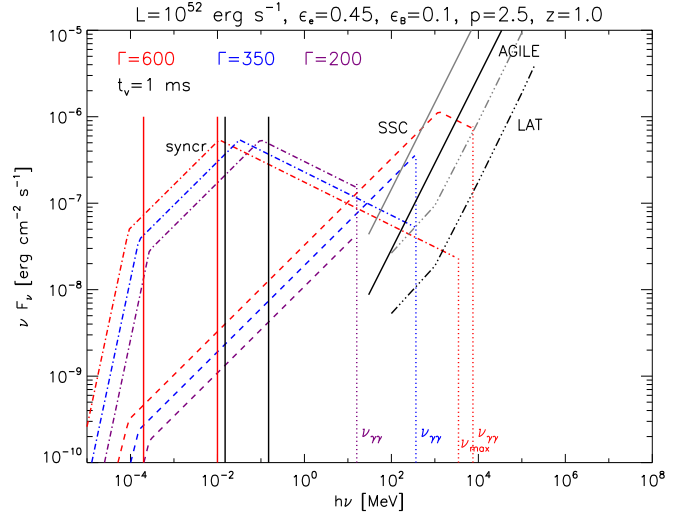}
\caption{Synchrotron (dot-dashed lines) and SSC (dashed lines) prompt emission spectra for a burst at redshift $z=1$. The burst prompt emission luminosity is $L_{52}$=1 and its variability is $t_v=1~ms$. We fix the fraction of fireball energy going into relativistic electrons and magnetic field to $\epsilon_e=0.45$ and $\epsilon_B=0.1$ respectively, and fix $p=2.5$. We display the predicted spectra for three different values of the fireball Lorentz factor $\Gamma$: 200 (purple), 350 (blue) and 650 (red). The solid lines and the dot-dot-dot-dashed lines represent AGILE and GLAST sensitivity for an integration time of 10 s (in grey) and 50 s (in black). The solid vertical lines refer to the \emph{Swift} XRT (red) and BAT (black) energy ranges.} 
\label{promptgammaz1}
\end{figure}

In Fig. \ref{promptgammaz1} we present the predicted synchrotron (dot-dashed lines) and SSC (dashed lines) prompt emission spectra as a function of the fireball Lorentz factor $\Gamma$ for a burst with luminosity $L=10^{52}~erg~s^{-1}$ located at redshift $z$=1. As in \citet{guetta03} we take $p$=2.5, $\epsilon_e=0.45$, $\epsilon_B=0.1$, $t_v=1$ ms and vary the fireball Lorentz factor $\Gamma$ between 200 and 600. In the same figure we report the AGILE (solid line) and GLAST (dot-dot-dot-dashed line) sensitivities (see Sect. \ref{detector} for details on the AGILE and GLAST sensitivity).
We note that the spectral cutoff energy due to pair production \citep[see Eq. (19) of][]{guetta03} moves to higher energies for higher fireball Lorentz factors while the peak of the prompt emission shifts to lower values. 
Therefore, given a set of parameters and a fixed integration time, the detectability of the SSC component improves with increasing fireball Lorentz factor values (see Fig. \ref{promptgammaz1}). The best candidate bursts for high energy detection will be those with the peak energy close to the lower value of the BAT energy band, $E_p \sim 10 $ keV. We find that the high energy emission for a burst with these characteristics will be detected both by AGILE and GLAST if $\Gamma>350 $ for an integration time $T_{int}=50~s$ and  $T_{int}=10~s$. In the most favorable case, $\Gamma\sim 600$ and $T_{int}=50~s$ , the SSC prompt emission component can be detected up to maximum redshift $z_{max} \sim 3.2$ by AGILE, and $z_{max} \sim 5.5$ by GLAST.


\begin{figure}[!htb]
\centering
\includegraphics[width=7.0cm,height=9.16cm,angle=90]{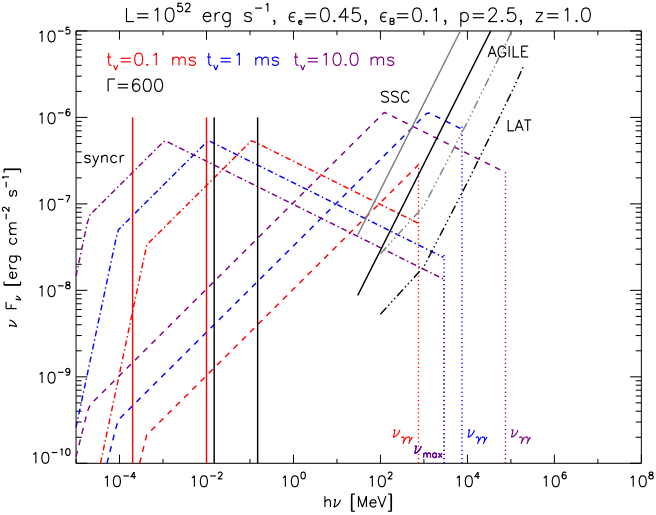}
\caption{Synchrotron (dot-dashed lines) and SSC (dashed lines) prompt emission spectra for a burst at redshift $z=1.0$. The fireball Lorentz factor is fixed to $\Gamma=600$, and the other parameters are the same as in Fig. \ref{promptgammaz1}. We display the predicted spectra for three different values of the burst temporal variability $t_v$: 0.1 ms (red), 1.0 ms (blue) and 10 ms (purple). The solid lines and the dot-dot-dot-dashed lines represent AGILE and GLAST sensitivity for an integration time of 10 s (in grey) and 50 s (in black). The solid vertical lines refer to the \emph{Swift} XRT (red) and BAT (black) energy ranges.} 
\label{prompttvz1}
\end{figure}

As in \citet{guetta03} we also study the predicted synchrotron and SSC emission as a function of the burst temporal variability $t_v$ for a burst at $z=1.0$ (Fig. \ref{prompttvz1}).  In this case, we fix $\Gamma=600$. The prompt SSC peak energy (Eq. \ref{piccoSSC}) moves to lower energy with larger $t_v$ while the spectral energy cutoff due to pair production \citep[Eq. 19 of][]{guetta03} moves to higher energies with larger $t_v$, therefore the detectability of the SSC prompt emission component improves with an increasing variability time $t_v$.

Figure \ref{prompttvz1} shows that the SSC component can be detected by AGILE and GLAST for $t_v=0.1-10$ ms both for 10 s and 50 s of integration time, and that the best condition for the detectability of the SSC component is achieved for $t_v=10~ms$. Therefore also in this case the best candidate bursts for high energy detection would be those with $E_p$ close to the lower value of the BAT energy band. For $t_v=10$ ms and an instrumental integration time $T_{int}=50$ s, the SSC prompt emission component can be detected up to maximum redshift $z_{max} \sim 6.5 $ by AGILE, and up to $z_{max} \sim 7.0$ by GLAST. We summarize our results in Table \ref{prompt}, where we report the maximum redshift at which AGILE and GLAST can detect a burst as a function of the fireball Lorentz factor $\Gamma$, the prompt emission temporal variability $t_v$ and the instrumental integration time $T_{int}$. Larger values of $\Gamma$ and $t_v$ shift the cutoff due to pair production to higher energies and the peak energy of the prompt emission to lower energy. Actually the peak energy of the synchrotron emission of these GeV emitters bursts falls below or inside the (15-150 keV) BAT band of Swift (see Fig. \ref{promptgammaz1}). For this purpose, the energy band covered by Swift  is better suited than the band covered by BATSE (300-500 keV) for the identification of the GeV burst sources.

\begin{table}[htb]
\caption{ Maximum redshift of detection by AGILE and GLAST as a function of the fireball Lorentz factor $\Gamma$, the prompt temporal variability $t_v$, and the instrumental integration time $T_{int}$. The other parameters are fixed as $L_{52}=1.0$, $\epsilon_e=0.45$, $\epsilon_B=0.1$, and $p=2.5$. We also report the peak energies $E_p$ of the synchrotron prompt emission and the cutoff energies $E_{cut}$ of the relative SSC component.}
\begin{center}
\label{prompt}
\begin{tabular}{c c c c c c c} 
\noalign{\smallskip} \hline \noalign{\smallskip} 
 satellite & $\Gamma$  &  $t_v$    & $T_{int}$ & $z_{max}$ & $E_p$  & $E_{cut}$  \\  
           &           &   [ms]    & [s]       &           & [keV]  &  [MeV]     \\
\hline \hline   
AGILE     & 600      &   1        & 10        & 1.8        &  7.78  & $5.4\times 10^3$ \\
\hline
 GLAST     & 600      &   1       & 10        & 3.0        &  5.44  & $3.8\times 10^3$ \\
\hline
 AGILE     & 600      &    1      & 50        & 3.2        &  5.18  & $3.6\times 10^3$ \\
\hline
 GLAST     & 600      &     1     & 50        & 5.5        &  3.35  & $2.3\times 10^3$ \\
\hline
 AGILE     &    600   &  10       & 10        & 3.0        &  0.54  & $3.8\times 10^4$ \\
\hline
 GLAST     &   600    &   10      & 10        & 4.0        &  0.44  & $3\times 10^4$ \\
\hline
 AGILE     &  600     &  10       & 50        & 6.5        &  0.29  & $2\times 10^4$ \\
\hline
 GLAST     &   600    &      10   & 50        & 7.0        &  0.27  & $1.9\times 10^4$ \\
\hline
 AGILE     &    200   &  10       & 10        & 1.5        & 7.8 & 127.9 \\
\hline
 GLAST     &   200    &   10      & 10        & 2.0        & 6.5 & 106.6 \\
\hline
 AGILE     &  200     &  10       & 50        & 2.0        & 6.5 &  106.6 \\
\hline
 GLAST     &   200    &  10       & 50        & 2.0        & 6.5 & 106.6 \\
\noalign{\smallskip}\hline

\end{tabular}
\end{center}
\end{table}

\section{X-ray and high energy flares in the context of internal shock}
\label{data}

As stated in Sect. \ref{intro}, several models have been proposed in the  literature to explain the origin of flares. Some GRBs show properties which can originate from LIS produced by a long lasting central engine activity \citep{chinca07,falcone07}. This motivated us to apply the IS model presented by \citet{guetta03} to the flare phenomenon and study the GeV emission from flares. In order to estimate the high energy counterpart of X-ray flares, we repeat the calculations given in the previous section.

\begin{figure}[!htb]
\centering
\includegraphics[width=7.0cm,height=9.16cm,angle=90]{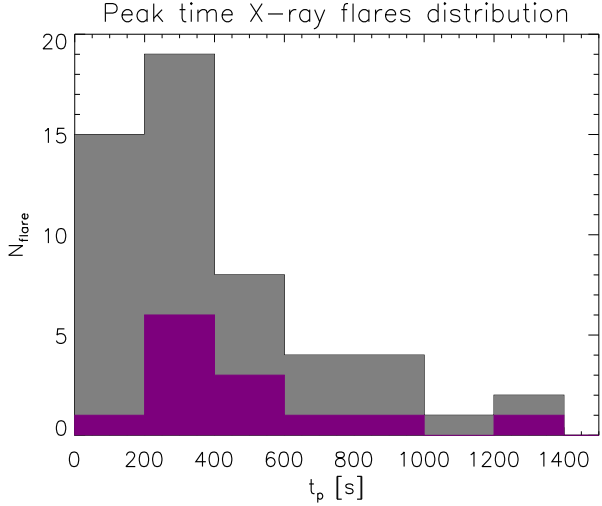}
\caption{Distribution of peak times of the total (with and without redshift) flare sample  of \citet{chinca07} (in grey), and of the sub-sample of flares with known redshift (in purple) reported in Table \ref{flare_properties}. }
\label{tempopicco}
\end{figure}

\begin{figure}[!htb]
\centering
\includegraphics[width=7.0cm,height=9.16cm,angle=90]{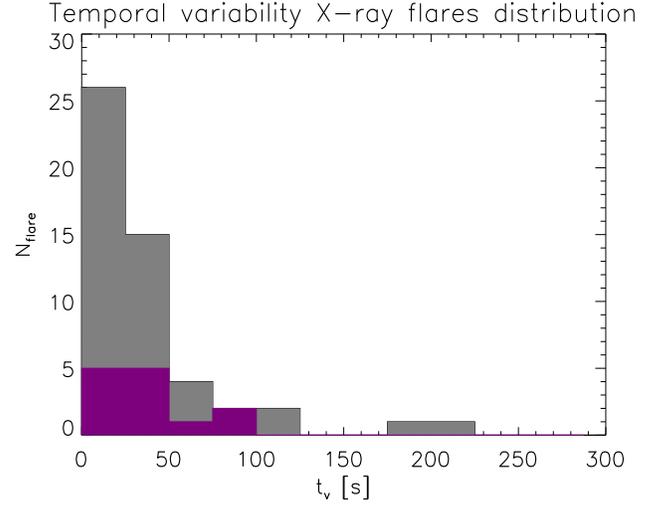}
\caption{Distribution of the flare duration of the complete sample of flare from \citet{chinca07} (in grey), and of the sub-sample of flares with known redshift reported in Table \ref{flare_properties} (in purple). }
\label{variability}
\end{figure}

\begin{figure}[!htb]
\centering
\includegraphics[width=7.0cm,height=9.16cm,angle=90]{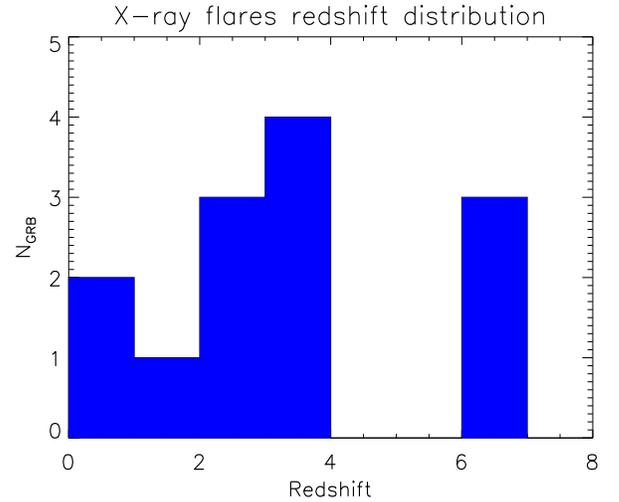}
\caption{Redshift distribution of the flares of our sample.}
\label{redshift}
\end{figure}

\begin{figure}[!htb]
\centering
\includegraphics[width=7.0cm,height=9.16cm,angle=90]{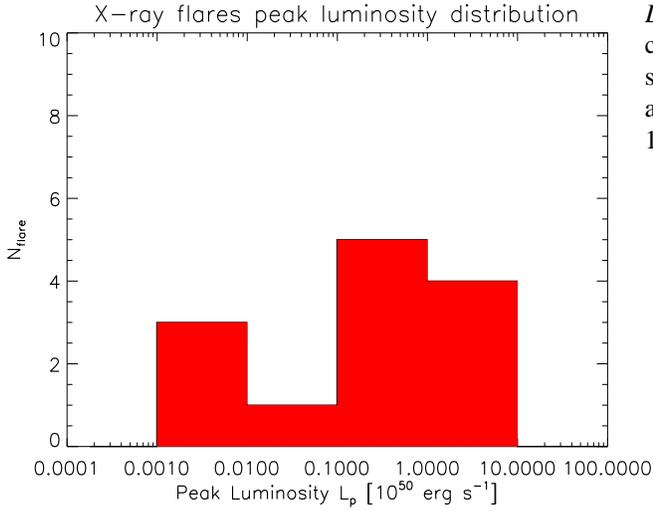}
\caption{Luminosity distribution of the flare of our sample. }
\label{luminosity}
\end{figure}

As shown in Sect. \ref{prompt}, in the LIS scenario the variability time $t_v$ (i.e. the average interval between consecutive shell ejections) has a fundamental role in the determination of the collision radii and consequently of all emission properties of X-ray flares. All the relevant quantities, such as the synchrotron (Eq. \ref{piccosincr}) and SSC (Eq. \ref{piccoSSC}) peak energy, with the corresponding fluxes and the high energy cutoff, depend strongly on $t_v$. If the mechanism responsible of the X-ray flares is the same as the prompt emission (IS) one could expect a similar variability time, thus we initially assume for flares that $t_v \sim$ 100 ms. However, given our ignorance on the average interval between consecutive shells when the engine is reactivated, that these shells could be emitted close to the central engine or at larger distances, and that the Lorentz factor $\Gamma$ could be smaller for X-ray flares in comparison with the prompt emission, in the following we also discuss the case of a larger $t_v$. The other relevant quantities are the luminosity $L$ of the flare and the average flare duration $t_f$. In order to determine the typical $t_f$ and $L$ of the flares, we select a sample of X-ray flares with known redshift and published light curves, taking place at times shorter than $\approx 1000$ s, when the IS mechanism can still be active. At longer times other mechanisms can be responsible for the flare emission, e.g. inhomogeneities in the external medium {\citep{lazzati07}. 

We take our sample of flares from \citet{chinca07}; in Table \ref{flare_properties} we report the flare properties which are relevant to our study.
\citet{chinca07} performed the temporal analysis of a sample of 33 GRBs detected by \emph{Swift} using a multi-broken power law to fit the underlying continuum and a Gaussian model for X-ray flares, and adopted the Gaussian width $\sigma$ as a measure of the X-ray flare duration $t_f$. In order to estimate the high energy emission which can be associated to the X-ray flares, we need the flare peak luminosity which we can get from the temporal and spectral information available in literature.

In Fig. \ref{tempopicco}, \ref{variability}, \ref{redshift} and \ref{luminosity} we report the flare peak time, duration, redshift and luminosity distributions of our sample of flares. For comparison we also report in the same figures the distributions of the complete (with and without known redshift) sample of flares presented by \citet{chinca07}. As we can see from these figures, the total sample of flare and our sub-sample of bursts with known redshift show the same distributions. For the total sample, the mean peak time of flares is $\langle t_p \rangle=404.0$ s, $\sigma=300.5$ and the mean duration is $\langle t_f \rangle= 40.4 s$, $\sigma=41.8$ s. We get similar values for the bursts of our sample, $\langle t_p \rangle=461.5$ s,  $\sigma =330.6$, and $\langle t_f \rangle= 39.6 s$, $\sigma=27.6$ \citep{chinca07} \citep[note that the fit of flares with two power laws gives flare durations of the order of hundreds of seconds, e.g.][]{falcone07}. We take the peak flux of each flare from published light curves and find their logarithmic mean peak luminosity, $\langle log L_p \rangle=49.3$, $\sigma= 1.2$ i.e $L_p \sim 2 \times 10^{49}~erg~s^{-1}$, while if we compute the ''linear'' medium we find $\langle L' \rangle \sim 10^{50}~erg~s^{-1}$. The mean redshift of the bursts of our sample is $\langle z \rangle=2.76$, $\sigma=1.92$.}

\begin{table*}[!htb]
\caption{Our sample of X-ray flares with known redshift and published light curves. $t_p$ is the time of the X-ray flare peak, $t_f$ is the flare duration, $\Delta t/t$ is the duration-to-flare time ratio, and $L_p$ is the peak luminosity of X-ray flares evaluated as $L_p= 4 \pi D_l^2 F_{0.2-10.0~KeV}(1+z)^{(2-\gamma)}$ with $\gamma$ the power law photon index. References: [1] \citet{romano06}; [2] \citet{chinca07}; [3] \citet{falcone07}; [4] \citet{perri07}; [5] \citet{cenko06} ; [6] \citet{cusumano07}; [7] \citet{oates06}.}
\centering
\label{flare_properties}
\begin{tabular}{c c c c c c c} 
\hline 
 name     & redshift & $t_p$ & {\bf$t_f$} & $\Delta t/t$ & $L_p$          & References \\  
          &    z     & [s]   & [s]   &              & $[erg~s^{-1}]$ &  \\
\hline \hline   
GRB050406 & 2.44     &$211 \pm 5$ & $17.9^{+12.3}_{-4.6}$ & 0.882 & $2.3\times10^{49}$ & [1] \\
\hline
GRB050724 & 0.26  & $275 \pm 5$ &  $30.6^{+6.6}_{-6.0}$ & - & $3.5 \times 10^{47}$ & [2],[3] \\
          &  -    & $327^{+6}_{-9}$   & $12.7^{+6.3}_{-5.0}$ & -& $1.6 \times 10^{47}$ & - \\
\hline
GRB050730 & 3.97     & $131.8^{12.7}_{-59.6}$ & $32.7^{+24.4}_{-8.3}$ & -  & $1.0\times10^{50}$ & [4]\\
          & -      & $234.2^{+2.7}_{-2.4}$ & $14.5^{+3}_{-2.8}$ & -  & $8.3\times10^{49}$  & - \\
          & -      & $436.5^{+1.5}_{-2.2}$ & $32.7^{+24.4}_{-8.3}$ & -  & $1.1\times10^{50}$ & -  \\
          & -      & $685.8^{+2.8}_{-2.7}$ & $23.8^{+3.9}_{-3.5}$ & -  & $7.8\times10^{49}$ & - \\
\hline
GRB050802 & 1.71    & $464 \pm 31$ & $100^{+33}_{-40}$ & 2.327 & $4.3 \times 10^{48}$ & [2],[3] \\
\hline
GRB050820A & 2.61    & $241^{+0}_{-1}$ & $9.5^{+0.3}_{-0.2}$ & -  & $1.2\times10^{50}$ & [5] \\
\hline
GRB050904 & 6.3      & $448.6^{+3.7}_{-4.0}$ & $45.9^{+4.5}_{-3.8}$ & -  & $8.1\times10^{50}$ & [6] \\
          & 6.3      & $975.5^{+38.5}_{-32.5}$ & $62.8^{+36.9}_{-32.5}$ & -  & $8.5\times10^{49}$ & - \\
          & 6.3      & $1265.5^{+28.0}_{-27.0}$ & $81.6^{+30.1}_{-28.2}$ & - & $8.7\times10^{49}$ & - \\
 \hline
GRB060108 & 2.03     & $303.5^{+23.5}_{-24.5}$ & $44.5^{+125.5}_{-30.5}$ & 1.405 & $4.0\times10^{47}$ & [7] \\
\hline
\end{tabular}
\end{table*}

To estimate the high energy counterpart of X-ray flares, we repeat the calculations made in Sect. \ref{prompt} by changing only the luminosity from $L=10^{52}~erg s^{-1}$ to $L=2 \times 10^{49}~erg~s^{-1}$. 

As noted by \citet{fan07} it is difficult to predict the expected SSC emission as we do not have a good estimate of the typical Lorentz factor of the electrons accelerated at the shock to produce the X-rays. However we can give some constraints on the Lorentz factor $\Gamma$ of the colliding shells.  If we assume that the X-ray flares are produced by accelerated electrons which cool very fast, the condition $\nu_c < \nu_m$ introduces an upper limit to the fireball Lorentz factor $\Gamma$. In particular, for typical flare luminosity and temporal variability  $t_v$=100 ms, and for $\epsilon_e=0.45$, $\epsilon_B=0.1$ and $p=2.5$, we find $\Gamma \lesssim 340$. In addition, in order to have SSC GeV flare emission we have to require that the spectral cutoff energy due to pair production is $\gtrsim 1~GeV$. With the above quantities, using Eq. 19 of \citet{guetta03}, we find $\Gamma \gtrsim$ 60. We thus take in the following $\Gamma=100$ and $\Gamma=300$ as two possible Lorentz factor values. The peak energy of synchrotron and SSC flare emission are given by:

\begin{equation}
\label{piccosincrflare}
E_{p,flare} = 3.8 (1+z)^{-1} (1+Y)^{-1/3} \epsilon_e^{3/2} \epsilon_B^{1/2} L_{49}^{1/2} \Gamma_{2}^{-2} t_{v,-1}^{-1}~keV
\end{equation}

and

\begin{equation}
\label{piccoSSCflare}
E_{p, flare}^{SC}= 1.5 (1+z)^{-1} \epsilon_e^{7/2}\epsilon_B^{1/2} L_{49}^{1/2} \Gamma_{2}^{-2} t_{v,-1}^{-1}~GeV
\end{equation}

with L$_{49}$ the flare peak luminosity in unity of $10^{49}~erg~s^{-1}$,  $\Gamma_{2}$ the flow Lorentz factor in unity of 100, and $t_{v,-1}$ the flare temporal variability in unity of 100 ms.  For typical parameters values the peak of flare synchrotron emission is inside or just below the XRT band and the peak of flare SSC emission is around ten-hundred MeV. 
The shells producing the flares collide at a radius $R \sim 6 \times 10^{13}$ cm from the central engine for $\Gamma=100$, i.e close to the typical distances where the collisions producing the prompt emission take place. This is an important result because, as stressed by \citet{fan07} the knowledge of the location of the shocks determines the parameters of the emission, which strongly affect the estimation of the SSC component.

\begin{figure}[!htb]
\centering
\includegraphics[width=7.0cm,height=9.16cm,angle=90]{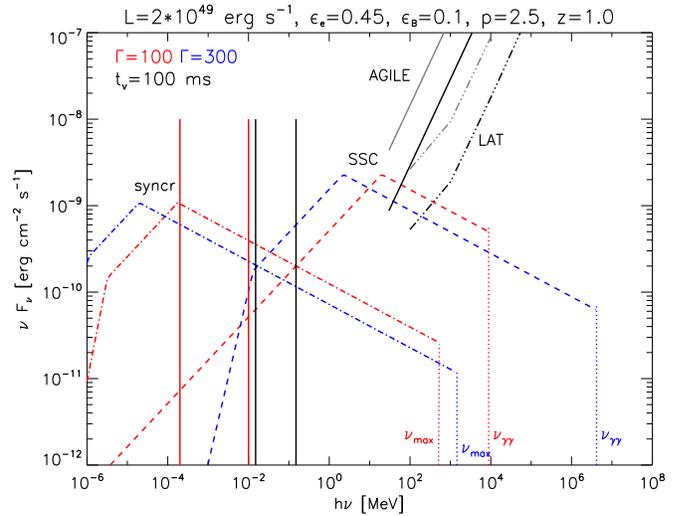}
\caption{Synchrotron (dot-dashed lines) and SSC (dashed lines) spectra for an X-ray flare with luminosity $L=2 \times 10^{49}~erg~s^{-1}$ and temporal variability $t_v=100$ ms at redshift $z=1$ as a function of the fireball Lorentz factor $\Gamma$. Other model parameters are the same as in Fig. \ref{promptgammaz1}. We display the predicted spectra for $\Gamma$=100 (red) and $\Gamma$=300 (blue). The solid and dot-dot-dot-dashed lines represent AGILE and GLAST sensitivity for an integration time of 100 s (in grey) and 500 s (in black). The solid vertical lines refer to the \emph{Swift} XRT (red) and BAT (black) energy ranges. }
\label{flarelum2e49z1}
\end{figure}

As for the prompt emission we study the predicted synchrotron and SSC flare emission as a function of the fireball Lorentz factor $\Gamma$, at redshift $z=1$ (Fig. \ref{flarelum2e49z1}). The flare duration is estimated to vary from tens to hundreds of seconds depending on the model used for the fit (i.e. a Gaussian or a power law respectively), we thus assume for AGILE and GLAST integration times of 100 s and 500 s. We find that for a flare luminosity $L=2 \times 10^{49}~erg~s^{-1}$ the (predicted) high energy flare emission can be detected by AGILE and GLAST up to a maximum redshift $z_{max} \sim 1.2$ and $z_{max} \sim 1.4$ for an integration time of 500 s, and up to $z_{max} \sim 0.6$ and $z_{max} \sim 0.7$ for an integration time of 100 s.

\begin{figure}[!htb]
\centering
\includegraphics[width=7.0cm,height=9.16cm,angle=90]{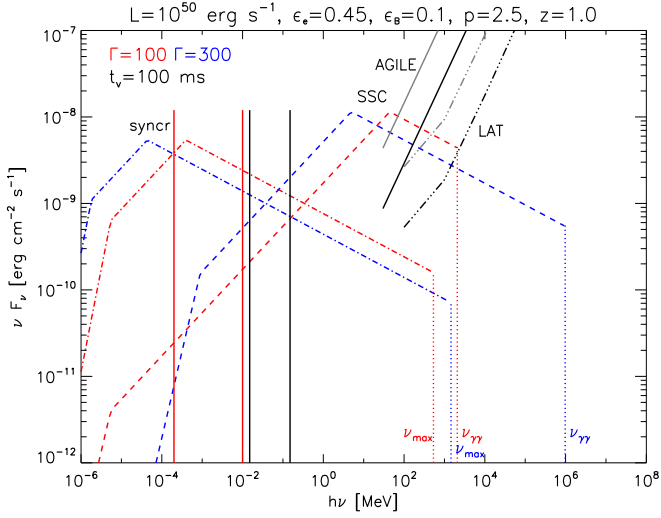}
\caption{Synchrotron (dot-dashed lines) and SSC (dashed lines) spectra for an X-ray flare with luminosity $L= 10^{50}~erg~s^{-1}$ and temporal variability $t_v=100$ ms at redshift $z=1$ as a function of the fireball Lorentz factor $\Gamma$. The other model parameters are the same of Fig. \ref{promptgammaz1}. We display the predicted spectra for $\Gamma$=100 (red) and $\Gamma$=300 (blue). The solid and dot-dot-dot-dashed lines represent AGILE and GLAST sensitivity for an integration time of 100 s (in grey) and 500 s (in black). The solid vertical lines refer to the \emph{Swift} XRT (red) and BAT (black) energy ranges. } 
\label{flarelum2e50z1}
\end{figure}

We then calculate the emission expected for X-ray flare luminosity $L'=10^{50}~erg~s^{-1}$ as a function of the fireball Lorentz factor $\Gamma$, at $z=1$ (Fig. \ref{flarelum2e50z1}). With this X-ray flares luminosity the predicted high energy emission can be detected both by AGILE and GLAST at $z=1$. If we take as mean flare luminosity $\langle L' \rangle = 10^{50}~erg~s^{-1}$, the SSC flare emission can be detected by AGILE and GLAST up to maximum redshift $z_{max} \sim 2.4$ and $z_{max} \sim 2.8$ for an integration time of 500 s, and up to $z_{max} \sim 1.2$ and $z_{max} \sim 1.4$ for an integration time of 100 s.

\begin{figure}[!htb]
\centering
\includegraphics[width=6.5cm,height=8.5cm,angle=90]{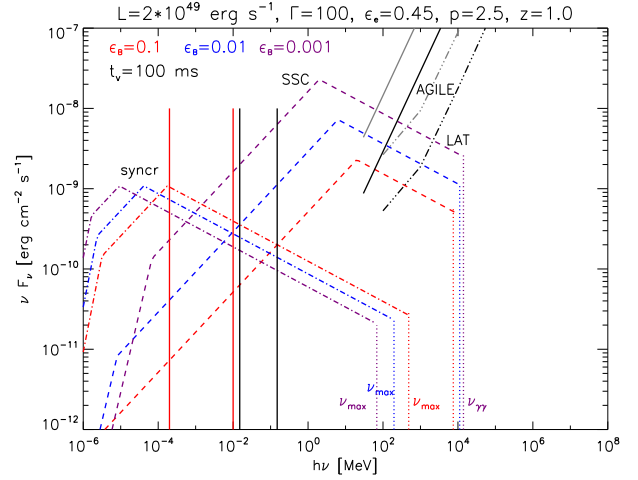}
\caption{Synchrotron (dot-dashed lines) and SSC (dashed) spectra for an X-ray flare with luminosity $L=2 \times 10^{49}~erg~s^{-1}$ and temporal variability $t_v=100$ ms at redshift $z=1$ as a function of $\epsilon_B$. The fireball Lorentz factor is fixed to $\Gamma=100$, and other model parameters are the same as in Fig. \ref{promptgammaz1}. We display the predicted spectra for $\epsilon_B$=0.1 (red), $\epsilon_B$=0.01 (blue) and $\epsilon_B$=0.001 (purple). The solid and dot-dot-dot-dashed lines represent the AGILE and GLAST sensitivities for an integration time of 100 s (in grey) and 500 s (in black). The solid vertical lines refer to the \emph{Swift} XRT (red) and BAT (black) energy ranges.} 
\label{flarelum2e49z1Eb}
\end{figure}

We also study how the relative importance of synchrotron and SSC emission varies with the ratio $(\epsilon_e/\epsilon_B)$, this ratio being the quantity which determines the relative importance of these two emission mechanisms \citep[for a detailed analysis of the effects of this ratio during early and late afterglow emission see][]{galli07}. We present our results in Fig. \ref{flarelum2e49z1Eb} for X-ray flare luminosity $L=2 \times 10^{49}~erg~s^{-1}$. As expected, the relative importance of the two processes increases with the ratio $(\epsilon_e/\epsilon_B)$ thus favoring the detection of the high energy component even though the peak of SSC emission moves toward lower energies with smaller $\epsilon_B$ values. We keep $\epsilon_e=0.54$ and vary $\epsilon_B$ from 0.1 (red curves), to 0.01 (blue curves), and finally to $0.001$ (purple curves). The predicted SSC emission can be detected by AGILE and GLAST up to $z_{max} \sim 2.5$ and $z_{max} \sim 2.7$ for an integration time of 500 s, and up to $z_{max} \sim 1.2$ and $z_{max} \sim 1.3$ for an integration time of 500 s. For a luminosity $L= 10^{50}~erg~s^{-1}$ the high energy emission can be detected by AGILE and GLAST up to $z_{max} \sim 5.2$ and $z_{max} \sim 5.5$ for an integration time of 500 s, and up to $z_{max} \sim 2.6$ and $z_{max} \sim 2.8$ for an integration time of 100 s.

\begin{figure}[!htb]
\centering
\includegraphics[width=6.5cm,height=8.5cm,angle=90]{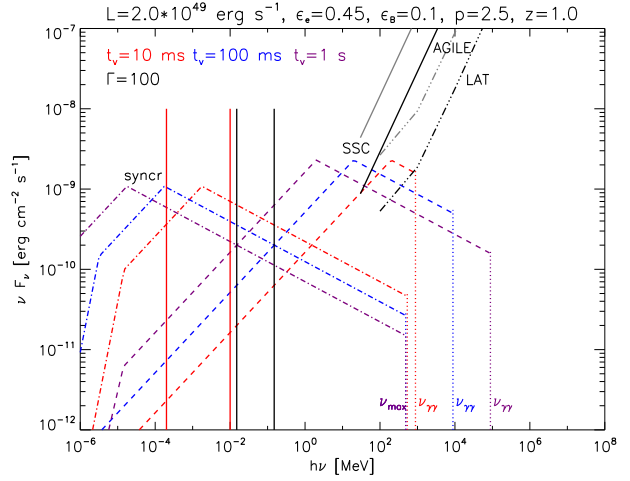}
\caption{Synchrotron (dot-dashed lines) and SSC (dashed) spectra for an X-ray flare with luminosity $L=2 \times 10^{49}~erg~s^{-1}$ as a function of the temporal variability $t_v$ at redshift $z=1$. The fireball Lorentz factor is fixed to $\Gamma=100$ and other model parameters are the same as in Fig. \ref{promptgammaz1}. We display the predicted spectra for $t_v$=10 ms (red), $t_v$=100 ms (blue) and $t_v$=1 s (purple). The solid and dot-dot-dot-dashed lines represent the AGILE and GLAST sensitivities for an integration time of 100 s (in grey) and 500 s (in black). The solid vertical lines refer to the \emph{Swift} XRT (red) and BAT (black) energy ranges.} 
\label{flarelum2e49z1tv}
\end{figure}

\citet{falcone07} have shown that X-ray flares spectra can be fitted both by a power law and/or by a Band model. However, due to the incomplete spectral coverage they could constrain the flare peak energy only in some cases, and suggested that typically this peak energy should be in the soft X-ray band, or between the optical and X-ray bands. Figures \ref{flarelum2e49z1}, \ref{flarelum2e50z1} and \ref{flarelum2e49z1Eb} show that if we assume $t_v\sim 100$ ms and $\Gamma \sim 100$, the peak energy of the X-ray flare emission falls in, or just below, the XRT band consistent with the findings of the flare spectral analysis performed by \citet{falcone07}. In Fig. \ref{flarelum2e49z1tv} we show the predicted synchrotron and SSC flare emission for a range of $t_v$ values similar to that typically observed during the prompt emission, i.e. $t_v \sim$ 10 ms - 1 s, with $\Gamma=100$ and $L=2 \times 10^{49}~erg~cm^{-2}$. As we can see from this figure, when one assumes a Lorentz factor of the order of one hundred for this range of $t_v$ the peak energy of the X-ray flare is always inside or just below the XRT band. The detection of the SSC component related to the X-ray flare depends strongly on the flare temporal variability: for $t_v <$ 10 ms the spectral cutoff due to pair production is shifted to lower energies \citep[Eq. 19 of][]{guetta03}, i.e. in the AGILE and GLAST bands, making the SSC emission more difficult to detect. In the same way, a larger temporal variability $t_v \gtrsim$ 1 s shifts the peak energy of the SSC component below the AGILE and GLAST energy bands, and this again makes the high energy flare more difficult to detect. In order to have both the X-ray peak energy in the XRT band and a detectable high energy emission for $\Gamma \sim$100, we need $t_v\sim 50-100$ ms. This implies that the flux of the X-ray flares should vary on a time scale much smaller than the flare duration, i.e $t_v<<t_f$. Therefore if LIS is the mechanism responsible for the flare emission, and the flare has a counterpart at high energies (from MeV to GeV), the X-ray flare light curve should present some substructures where the flux can even double its value at a time $t_v$ which is much smaller than the flare duration as observed in the prompt emission. 

We should note however that the temporal resolution of the \emph{Swift} XRT is $\sim$ 2 ms in WT mode and 2.6 s in PC mode (see http://www.swift.psu.edu/xrt/software.html\#modes). At the time of flare occurrence the readout is typically in the PC mode. Data are taken in WT mode only for the brightest flares, however the search of substructures with duration $\lesssim$ 1 s would be important to check the LIS model. If no significant variation in the X-ray flux on a small time scale is detected, then the average time interval $t_v$ between consecutive shell ejections should be the average value of the flare duration. For our sample we obtain $t_v \sim 40$ s, which would change the lower and upper limits of the flow Lorentz factor $\Gamma$ to $\sim$ 25 and $\sim$ 110 respectively. 

\begin{figure}[!htb]
\centering
\includegraphics[width=7.0cm,height=9.16cm,angle=90]{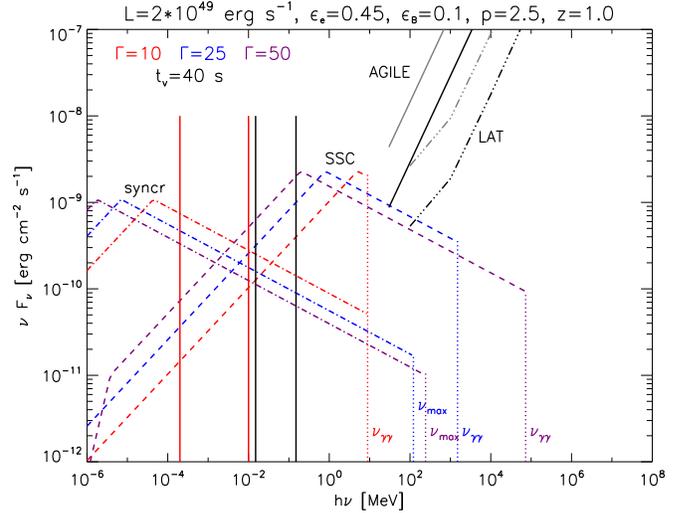}
\caption{Synchrotron (dot-dashed lines) and SSC (dashed lines) spectra for a typical X-ray flare with luminosity $L=2 \times 10^{49}~erg~s^{-1}$ and temporal variability $t_v=40~s$ at redshift $z=1$ as a function of the fireball Lorentz factor $\Gamma$. Other model parameters are the same as in Fig. \ref{promptgammaz1}. We display the predicted spectra for $\Gamma$=10 (red), $\Gamma$=25 (blue) and $\Gamma$=50 (purple). The solid and dot-dot-dot-dashed lines represent AGILE and GLAST sensitivity for an integration time of 100 s (in grey) and 500 s (in black). The solid vertical lines refer to the \emph{Swift} XRT (red) and BAT (black) energy ranges. }
\label{flarelum2e49z1tv40s}
\end{figure}

If one assumes the Lorentz factor to be of the order of 50, the LIS model predicts that the peak energy of the X-ray flare emission is around 1 eV, i.e. well below the XRT band (see Fig. \ref{flarelum2e49z1tv40s}). For $t_v=t_f=40$ s, $L=2 \times 10^{49}~erg~cm^{-2}$ and $\Gamma$=50, the shells producing the flare would collide at a radius $R \sim 6 \times 10^{15}$ cm from the central engine, i.e. at distances larger than that where the collisions producing the prompt emission take place. For $t_v$=40 s if one assumes a lower Lorentz factor as $\Gamma$=25 or $\Gamma=$10, then the X-ray flare peak energy moves to higher values consistent with those found for $t_v \sim$ 100 ms. Under these assumptions large temporal variability values could also account for the observed flare properties. Moreover, low values of the Lorentz factor as $\Gamma \sim$ 10 have been used by several authors, e.g. \citet{falcone06}, to fit X-ray flares. However in the case of low Lorentz factor values the spectral cutoff due to pair production goes below 1 GeV, and X-ray flares are not expected to have a high energy counterpart. Simultaneous observations by \emph{Swift} in the X-ray band, and by AGILE and GLAST at high energies, and a detailed temporal analysis of flares light curves are thus very important to test the LIS model we propose in this paper.

We summarize our redshift study for SSC flare emission in Table \ref{flares}, where we give the maximum redshift achievable by AGILE and GLAST for an integration time $T_{int}=100~s$ and $T_{int}=500~s$ as a function of the fireball Lorentz factor $\Gamma$, the flare luminosity $L$, the ratio $(\epsilon_e/\epsilon_B)$, and the flare temporal variability $t_v$. The peak energy of flares moves to higher energies with increasing flare peak luminosities and lower values of the $(\epsilon_e/\epsilon_B)$ ratio, while it shifts to lower energies for larger flow Lorentz factors. 

As can be seen from Table \ref{flares}, if LIS are responsible for X-ray flares and flare light curves contain some features where the flux can double its value at $t_v=10-100$ ms, similar to which is observed during the prompt emission, the peak energy of X-ray flares is in, or just below, the XRT energy band consistent with the spectral analysis performed by \citet{falcone07}. For $t_v=40$ s the peak energy is expected to be between the optical and the X-ray band only for low Lorentz factor values.

\begin{table*}[htb]
\caption{Maximum redshift of detection by AGILE and GLAST as a function of the fireball Lorentz factor $\Gamma$, the flare luminosity $L$, the ratio $(\epsilon_e/\epsilon_B)$, and the flare temporal variability $t_v$, and the instrumental integration time $T_{int}$.}
\centering
\label{flares}
\begin{tabular}{c c c c c c c c c} 
\hline 
satellite & $\Gamma$  &  $L$      & $\epsilon_e/\epsilon_B$ & $t_v$  & $T_{int}$ & $z_{max}$ & $E_p$     & $E_{cut}$\\ 
     &       &  $ [erg s^{-1}]$   &                         &  [ms]  &        [s]      &           & [keV]     & [MeV]   \\
\hline \hline   
 AGILE  & 100 & $2 \times 10^{49}$ &   4.5                  & 100    &       500       & 1.2       & 0.16      & $8.0 \times 10^{3}$ \\
\hline
 GLAST  & 100 & $2 \times 10^{49}$ &   4.5                  & 100    &        500      & 1.4       & 0.15      & $7.4 \times 10^{3}$ \\
\hline
 AGILE  & 100 & $2 \times 10^{49}$ &   4.5                  & 100    &     100         & 0.6       & 0.22      & $1.1 \times 10^{4}$ \\
\hline
 GLAST  & 100 & $2 \times 10^{49}$ &   4.5                  & 100    &      100        &  0.7      & 0.22      & $10^{4}$ \\
\hline
 AGILE  & 100 & $10^{50}$          &   4.5                  & 100    &        500      & 2.4       & 0.23      & $1.2 \times 10^{3}$ \\
\hline
 GLAST  & 100 & $10^{50}$          &   4.5                  & 100    &       500       & 2.8       & 0.21      & $1.1 \times 10^{3}$ \\
\hline
 AGILE  & 100 & $10^{50}$          &   4.5                  & 100    &     100         &  1.2      &  0.36     & $1.9 \times 10^{3}$ \\
\hline
 GLAST  & 100 & $10^{50}$          &   4.5                  & 100    &     100         & 1.4       &  0.33     &  $1.7 \times 10^{3}$ \\
\hline
 AGILE  & 100 & $2 \times 10^{49}$ &  450              & 100    &      500        & 2.5       & $5.2 \times 10^{-3}$ & $8.0 \times 10^{3}$ \\
\hline
 GLAST  & 100 & $2 \times 10^{49}$ &  450              & 100    &      500        & 2.7       & $4.9 \times 10^{-3}$ & $7.6 \times 10^{3}$ \\
\hline
AGILE  & 100 & $2 \times 10^{49}$ &  450              & 100     &   100           &  1.2      & $8.3 \times 10^{-3}$ & $1.3 \times 10^{4}$ \\
\hline
GLAST  & 100 & $2 \times 10^{49}$ &  450           & 100    & 100  &  1.3 & $8.0 \times 10^{-3}$ & $1.2 \times 10^{4}$ \\
\hline
AGILE  & 100 & $10^{50}$          &  450           & 100    &     500   & 5.2       & $6.6 \times 10^{-3}$ & $1.1 \times 10^{3}$ \\
\hline
GLAST  & 100 & $10^{50}$          & 450            & 100    &     500   & 5.5       & $6.3 \times 10^{-3}$ &  $10^{3}$  \\
\hline 
AGILE  & 100 & $10^{50}$          &  450           & 100    &  100 &  2.6 & $1.1 \times 10^{-2}$ & $1.8 \times 10^{3}$ \\
\hline
GLAST  & 100 & $10^{50}$          & 450            & 100    &  100 &  2.8 & $1.1 \times 10^{-2}$ & $1.7 \times 10^{3}$ \\
\hline 
AGILE  &  25 & $2 \times 10^{49}$ & 4.5   & $4 \times 10^4$ &     500   &  0.8 & $7.8 \times 10^{-3}$ & $1.7 \times 10^{3}$ \\
\hline
GLAST  &  25 & $2 \times 10^{49}$ & 4.5   & $4 \times 10^4$ &     500   & 0.9 & $7.4 \times 10^{-3}$ & $1.6 \times 10^{3}$ \\
\hline
AGILE  &  25 & $2 \times 10^{49}$ & 4.5   & $4 \times 10^4$ &  100 & 0.35 &  0.01 & $2.2 \times 10^{3}$ \\
\hline
GLAST  & 25 & $2 \times 10^{49}$ & 4.5   & $4 \times 10^4$ &  100 &  0.4  & 0.01 & $2.2 \times 10^{3}$ \\
\hline
\end{tabular}
\end{table*}

\section{Summary and conclusions}
\label{conclusions}

We have calculated the synchrotron and SSC emission during the prompt and the X-ray flare GRB phases from IS and studied how they depend on model parameters. For $p=2.5$ (the power law index of the electron energy distribution) and typical model parameters, the SSC component dominates the prompt emission above $\approx 100 $ MeV and the flare emission above $\approx 100$ keV. As can be seen from Eq. 19 of \citet{guetta03} and Eq. \ref{piccosincr} and \ref{piccoSSC}, and from Fig.  \ref{promptgammaz1} and \ref{prompttvz1}, during the prompt phase larger values of the flow Lorentz factor $\Gamma$ and the temporal variability $t_v$ shift the cutoff energy to higher values, and the peak of the emission to lower values. For example, in order to have $\sim 1$ GeV emission for $t_v=1$ ms, we need $\Gamma>350$ which implies $E_p<100$ keV exactly in the range of BAT (Swift).  Therefore we expect that a good fraction of the bursts detected by Swift can produce GeV emission which will be detected by AGILE and GLAST. 
  
We have taken the LIS as the mechanism responsible for the flare emission. In this model the average interval between consecutive shell ejection $t_v$, has a fundamental role in determining the radius of the collision and therefore the emission properties (i.e. the peak energy) of X-ray flares. We initially considered a range of $t_v$ (10 ms-1 s) and a Lorentz factor ($\Gamma \sim$100) similar to the prompt emission, and found that the peak of the emission is between the optical and the X-ray bands, consistent with the observations. This implies that the X-ray flare light curve should contain some substructures of $t_v\sim 100$ ms such as the prompt emission. An accurate search of these flux variations is mandatory to understand the physics of the flares and to check the LIS model. If no flux variation on a small time scale is detected the implication would be that the shells are ejected with an average interval similar to the duration of the single flare ($t_f\sim 40 s$). In this case the collisions between the shells happen at a large distance from the central engine. \citet{spada00} have shown that the pulse duration increases with the distance $R$ from the central engine, and consequently with time. This is consistent with the observations which show that the flare duration increases with the time of flare appearance \citep{chinca07}. If $t_v=t_f \sim$ 40 s the peak energy of the X-ray flare emission is between the optical and the X-ray band only for low Lorentz factors, $\Gamma \sim 10 \div 25$. However, in this case the cutoff due to pair production is below 1 GeV and X-ray flares are expected to not have a counterpart at high energies.

We have studied the detectability of the synchrotron and SSC emission during the prompt and the X-ray flare GRB phases as a function of the model parameters. During the GRB prompt phase we find that the detectability of the SSC emission component improves with increasing fireball Lorentz factor values $\Gamma$ and increasing $t_v$. For a typical prompt luminosity $L=10^{52}~erg~s^{-1}$ the SSC prompt emission component can be detected by AGILE and GLAST up to a maximum redshift $z_{max} \sim 3.0$ and $z_{max} \sim 4.0$ for an integration time $T_{int}=10~s$, and up to $z_{max} \sim 6.5 $ and $z_{max} \sim 7.0$ for an integration time $T_{int}=50~s$.

During the X-ray flare phase the detectability of the SSC component improves with smaller values of the fireball Lorentz factor $\Gamma$ depending on the position of the SSC peak energy and the cutoff energy due to pair production with respect to the observational band of the detector, and with larger values of the $(\epsilon_e/\epsilon_B)$ ratio. For a flare luminosity $L=2 \times 10^{49}~erg~s^{-1}$ and $t_v=100$ ms the SSC component can be detected by AGILE and GLAST up to a maximum redshift $z_{max} \sim 2.5$ and $z_{max} \sim 2.7$ for an integration time of 500 s, and up to $z_{max} \sim 1.2$ and $z_{max} \sim 1.3$ for an integration time of 100 s. If we assume a more optimistic mean flare luminosity $L=2 \times 10^{50}~erg~s^{-1}$ the SSC component can be detected by AGILE and GLAST up to a maximum redshift $z_{max} \sim 5.2$ and $z_{max} \sim 5.5$ for an integration time of 500 s, and up to and up to $z_{max} \sim 2.6$ and $z_{max} \sim 2.8$ for an integration time of 100 s.

Our findings show that AGILE and GLAST have good possibilities for detecting the high energy emission coming from flares. In particular, a comparative study of X-ray synchrotron emission and high energy SSC emission is an important tool to test the X-ray flare model we present in this paper, and/or to disentangle between the several models proposed in the literature to explain the origin of flares. In the context of LIS models, an additional mechanism producing high energy flares is possible (with respect to the SSC process considered in this paper), i.e the external inverse Compton scattering of the X-ray flare photons on the afterglow electrons. We can discriminate this component from the SSC, because in this case we expect that the high energy flares last much longer than the X-ray flare \citep{wang06,fan07}. A distinctive element between the variant of the IS model which we study in detail in this paper and models in which flares are caused by an ES \citep[e.g][]{galli07} is that in the context of LIS the peak of the SSC flare component is expected to be at lower energies with respect to that of the ES SSC component, i.e. in the MeV band for IS and in the GeV-TeV band for ES.

\begin{acknowledgements}
The authors are grateful to L. Stella for useful comments and discussions.
\end{acknowledgements}

\end{document}